\documentclass[11pt]{article}
\usepackage{amsmath,amssymb,amsthm,mathrsfs,stackrel}
\usepackage{amsfonts,bm,amsbsy,bbm}
\usepackage{rotating,lscape}
\usepackage{array}
\usepackage{graphicx,psfrag,epsf}
\usepackage{multirow}
\usepackage{multicol}
\usepackage{threeparttable}
\usepackage{booktabs}
\usepackage[round]{natbib}
\usepackage{float}
\usepackage{algorithm}
\usepackage{algpseudocode}
\usepackage{url}
\setcitestyle{authoryear,round}
\usepackage[colorlinks,citecolor=blue,linkcolor=blue,urlcolor=blue]{hyperref}
\usepackage{comment}
\usepackage{tikz-cd}

\newtheorem{Rem}{Remark}
\newtheorem{Assum}{Assumption}
\newtheorem{Prop}{Proposition}
\newtheorem{lemma}{Lemma}
\newtheorem{theorem}{Theorem}

\def\bX{\boldsymbol{X}}
\def\bW{\boldsymbol{W}}
\def\bZ{\boldsymbol{Z}}
\def\bz{\boldsymbol{z}}

\def\bp{\boldsymbol{p}}

\def\bh{\boldsymbol{h}}
\def\bv{\boldsymbol{v}}
\def\cA{\mathcal{A}}
\def\bgamma{\boldsymbol{\gamma}}

\def\btheta{{\boldsymbol{\theta}}}

\def\bmu{\boldsymbol{\mu}}

\def\mR{\mathbb R}

\begin{document}

\title{\textbf{Simulation-free extrapolation for misspecified models induced by categorizing an error-prone continuous covariate}}

\author{
Huali Zhao\textsuperscript{1} and 
Tianying Wang\textsuperscript{2}\thanks{Tianying.Wang@colostate.edu}\\[0.5em]
\textsuperscript{1}Department of Mathematical Sciences, Tsinghua University\\
\textsuperscript{2}Department of Statistics, Colorado State University
}

\date{}

\maketitle
\thispagestyle{empty}
\baselineskip=20pt

\begin{abstract}
Epidemiological studies often categorize continuous exposures for interpretation even when the underlying outcome-exposure association is continuous. The fitted categorical regression is a misspecified model because it replaces the continuous exposure with categories. With measurement error, categorization also misclassifies latent exposure categories, so the observed regression generally targets different means and contrasts. Existing estimating-equation and simulation-based extrapolation approaches require, respectively, outcome-model-specific derivations and pseudo-data generation with repeated fitting. We introduce simulation-free extrapolation (\mbox{SIMFEX}), which estimates misclassification probabilities and latent category proportions from replicates, computes the mean-scale trajectory without pseudo-data or repeated outcome-model fitting, and extrapolates it to the no-misclassification endpoint. Without additional error-free covariates, this construction applies across known one-to-one links, and the resulting estimator is consistent under stated conditions. With error-free covariates, the contrast relation remains exact for identity-link additive models when misclassification probabilities and category proportions are covariate-invariant; the nonidentity-link version provides a practical approximation. Simulations show substantial bias reduction relative to the naive analysis and coverage generally close to nominal. In the UK Biobank analysis, SIMFEX produced larger estimated high-versus-low fat-intake contrasts than the naive analysis for body mass index and obesity, illustrating how category misclassification can change the magnitude and uncertainty of prespecified contrasts.
\end{abstract}

\noindent
\textbf{Keywords:} categorized exposure; measurement error; misclassification; model misspecification; simulation extrapolation.

\pagestyle{plain}

\begingroup
\allowdisplaybreaks
\section{Introduction}\label{sec:introduction}
Nutrient intake plays a central role in studies of many health outcomes, and epidemiological studies commonly report associations by comparing categories of continuous dietary exposures \citep{nci,arem2013healthy,jessri2023mortality}. Such comparisons are readily interpretable and align with scientific questions about how disease risk differs across intake levels. At the same time, nutrient intake is often measured through self-reported dietary instruments and is therefore subject to well-recognized measurement error \citep{brown1994energy,achic2018categorizing}. When an error-prone continuous exposure is categorized, measurement error can move individuals across category cutpoints and change the resulting category comparisons. We develop a correction that preserves the familiar category-based analysis while accounting for measurement error in the underlying continuous exposure.

\subsection{Misspecified model in epidemiological studies}
In epidemiological practice, investigators often study an association that varies over a continuous exposure scale but report results from a categorical regression because its contrasts have direct scientific interpretations. For example, \citet{reedy2008index} categorized continuous dietary measures, including saturated fat, polyunsaturated fat, and alcohol intake, into quintiles and compared disease risk across intake groups. Highest-versus-lowest category comparisons are also common in other nutritional epidemiology studies \citep{arem2013healthy,jessri2023mortality}.

Following  \citet{achic2018categorizing}, we call the fitted categorical regression a \emph{misspecified model} because the underlying exposure-response association is defined on the continuous exposure scale, whereas the empirical analysis replaces that exposure by its categorized version. The terminology refers to this difference in scale between the underlying association and the fitted empirical model. The category-specific outcome means and their contrasts remain well-defined and scientifically meaningful. Accordingly, we treat the categorical misspecified model as the primary empirical analysis and its category-specific means and contrasts as the scientific targets.

Measurement error creates a distinct problem. Let the intended categories be defined by the latent continuous exposure $X$, while the observed analysis categorizes an error-prone measurement $W$ using the same cutpoints. The observed categories then combine individuals from different true exposure categories, so the fitted analysis generally estimates different category-specific means and contrasts. The statistical task is therefore to recover the parameters of the intended categorical misspecified model from the category misclassification induced by measurement error \citep{gustafson2002comparing,achic2018categorizing}. This is the correction problem addressed by SIMFEX.

\subsection{Measurement error in 24-hour dietary recall questionnaires}\label{subsec:data description}
The 24-hour dietary recall data from the UK Biobank provide a concrete example. Participants reported the foods and beverages consumed during the preceding 24 hours on two occasions, from which daily nutrient intakes were calculated. A short-term recall varies around the latent long-term usual intake, and the repeated measurements reveal the resulting within-person variability. When usual intake is the exposure of interest, categorizing a daily recall can therefore assign an individual to a different intake category.

The left panel of Figure~\ref{fig:intro} shows the estimated misclassification matrix for quintiles of daily fat intake, with estimation details given in Section~\ref{sec:data analysis}. The first and fifth categories have estimated misclassification rates of approximately 37\%, while the estimated rates for the three middle categories exceed 70\%. These movements across category cutpoints produce substantial mixing of the exposure groups underlying the intended comparisons. The right panel shows that observed daily fat intake departs from a Gaussian reference distribution. This category mixing motivates the category-level correction, whereas the distributional departure motivates modeling the continuous exposure distribution when estimating the induced misclassification from replicate information.
\begin{figure}[!ht]
    \centering
    \includegraphics[scale=0.55]{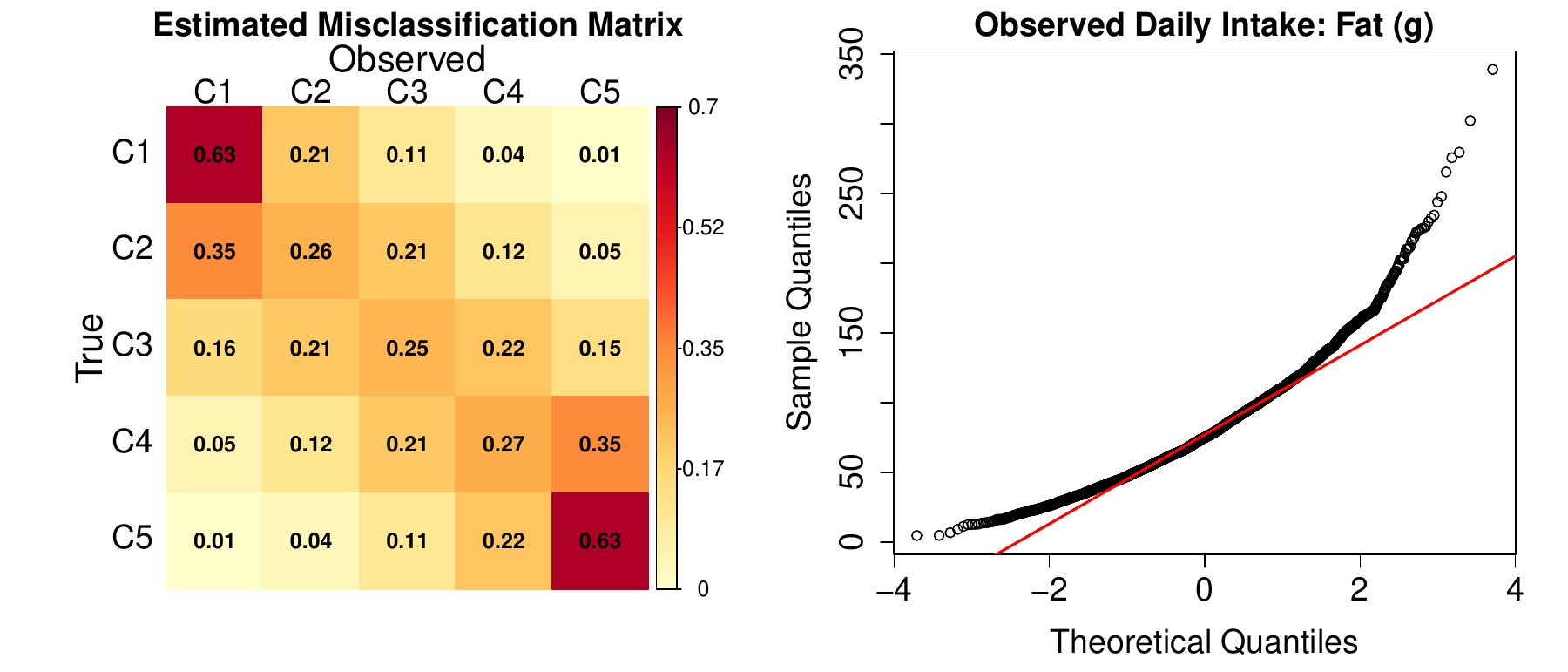}
    \caption{Misclassification and distributional features of daily fat intake in the UK Biobank data. Left: estimated misclassification probabilities for quintiles of daily fat intake. Right: quantile-quantile (QQ) plot of observed daily fat intake.}
    \label{fig:intro}
\end{figure}

Two correction strategies address this problem through different constructions. The estimating-equation method of \citet{achic2018categorizing} derives an observed-data equation from a specified outcome model on the continuous exposure scale, so a new equation is needed when the outcome model changes. MCSIMEX \citep{kuchenhoff2006general} operates directly on a misclassified categorical predictor but generates pseudo-datasets and repeatedly refits the outcome model along the extrapolation path. These considerations motivate a category-level correction that estimates the misclassification quantities from replicates and computes the extrapolation trajectory without pseudo-data generation or repeated fitting. In the baseline setting without additional error-free covariates, an additional objective is to reuse the same downstream correction across outcome models without a new model-specific derivation.

\subsection{Our contribution}
Our principal result is a population mean-scale relation that links outcome means defined by the observed categories to those defined by the latent exposure categories through the category-misclassification mechanism and the true category probabilities. This relation leads to \textbf{SIM}ulation-\textbf{F}ree \textbf{EX}trapolation, or SIMFEX, which uses replicate-estimated nuisance quantities and deterministic extrapolation to estimate the parameters of the intended categorical misspecified model. Our main contributions are as follows:
\begin{enumerate}
    \item The population relation expresses the observed-category outcome means as a transformation of the error-free category means through the misclassification matrix and the true category probabilities. In this baseline setting, constructing the correction on the mean scale allows the same formulation to accommodate identity, logistic, probit, and other known one-to-one links without a separate correction for each outcome model.
    
    \item This relation yields a simulation-free estimator that computes the pseudo-mean trajectory deterministically and extrapolates it to the formal error-free endpoint. In contrast to MCSIMEX \citep{kuchenhoff2006general}, SIMFEX does not generate misclassified pseudo-datasets or repeatedly refit the outcome model.

    \item We provide an estimable construction for the misclassification matrix and the true category probabilities using replicate measurements of the continuous exposure. This connects the continuous measurement error information to the category-level correction, and we establish consistency of the resulting SIMFEX estimator under the stated conditions.

    \item We extend SIMFEX to covariate-adjusted analyses. For identity-link additive models, we derive an exact population contrast relation when the category-misclassification mechanism and true category probabilities do not vary with the error-free covariates, and use it to construct the corresponding deterministic extrapolation estimator. For nonidentity links, the same deterministic strategy provides a practical approximation whose finite-sample behavior is examined numerically.
\end{enumerate}

The rest of the paper is organized as follows. Section~\ref{sec:background} reviews related measurement error methods and defines the target categorical misspecified model. Section~\ref{sec:methodology} develops SIMFEX and its covariate-adjusted extension, Sections~\ref{sec:simulation} and \ref{sec:data analysis} present simulation studies and a UK Biobank application, and Section~\ref{sec:conclusion} concludes with a discussion.

\section{Background}\label{sec:background}
\subsection{Literature review on measurement error analysis}\label{subsec:measurementerror}
Classical approaches to covariate measurement error include simulation-based, estimating-equation, and likelihood methods. SIMEX \citep{stefanski1995simulation} generates pseudo-data with increasing measurement error and extrapolates the resulting estimator trajectory to the no-error case, while MCSIMEX adapts this principle to misclassified covariates \citep{kuchenhoff2006general}. Estimating-equation and likelihood approaches incorporate information about the latent covariate and the measurement-error mechanism through regression calibration, corrected estimating equations, or likelihood-based calculations \citep{schafer1993likelihood,schafer1996likelihood,novick2002corrected,kipnis2012regression}. These methods provide general routes for measurement-error correction. The problem considered here additionally requires connecting continuous-scale measurement-error information to the categorical predictor used in the fitted regression model. Appendix~1 gives broader context on measurement error, endogeneity, and causal estimands. 

When an error-prone continuous exposure is categorized, measurement error induces misclassification of the predictor used in the fitted model. \citet{gustafson2002comparing} studied bias correction when an underlying continuous predictor is measured with error and then dichotomized. \citet{achic2018categorizing} developed an estimating-equation treatment for regression parameters associated with a categorized continuous predictor. Their construction begins with a continuous-scale response model and forms an observed-data estimating equation through conditional expectations involving the latent predictor $X$, so the equation is derived for the selected response and measurement-error models. SIMFEX instead uses replicate measurements to estimate the category-level misclassification quantities and then performs the outcome correction through a deterministic calculation on the category-mean scale. This separation yields a common downstream correction for known one-to-one links in the baseline formulation. Section~\ref{subsec:connection} gives a detailed comparison of these approaches.

\subsection{Target parameter and notation}\label{subsec:notationmodel}
Let $Y$ denote the response, which may be continuous or binary, and let $X\in\mR$ denote the underlying continuous exposure. Let $(C_1,\ldots,C_J)$ be a prespecified partition of the exposure scale, chosen according to the scientific aim of the study, such as tertiles ($J=3$) or quintiles ($J=5$). Define the categorized exposure vector
\[
\bX^c
=
(X^c_1,\ldots,X^c_J)^\top
=
(I_{X\in C_1},\ldots,I_{X\in C_J})^\top.
\]
Here, $I_A$ denotes the indicator of event $A$.

For a known one-to-one link function $g$ on the relevant mean domain, define the error-free category-specific target parameter $\btheta=(\theta_1,\ldots,\theta_J)^\top$ through the categorical misspecified model
\begin{equation}\label{eq:misspecifiedmodel}
    g\{E(Y\mid \bX^c)\}=\btheta^\top\bX^c.
\end{equation}
At the category level, model~\eqref{eq:misspecifiedmodel} is saturated, so
\[
g^{-1}(\theta_j)=E(Y\mid X^c_j=1),
\qquad
j=1,\ldots,J.
\]
Thus, $\btheta$ is defined directly by the category-specific means without requiring a parametric model for the continuous-scale regression $E(Y\mid X=x)$. Following \citet{achic2018categorizing}, we use the term \emph{misspecified model} because the underlying exposure-response association is defined on the continuous exposure scale, whereas the empirical analysis uses the categorized exposure $\bX^c$. Because the components of $\bX^c$ sum to one, no additional intercept is included; Appendix~2.1 gives further identifiability details.

When $X$ is unobserved, let $W$ denote its error-prone measurement and categorize $W$ using the same cutpoints:
\[
\bW^c
=
(W^c_1,\ldots,W^c_J)^\top
=
(I_{W\in C_1},\ldots,I_{W\in C_J})^\top.
\]
The \emph{naive} analysis substitutes $\bW^c$ for $\bX^c$ in model~\eqref{eq:misspecifiedmodel} and fits
\begin{equation}\label{eq:misspecified_model_error}
    g\{E(Y\mid \bW^c)\}=\btheta_{\emph{naive}}^\top\bW^c,
\end{equation}
where $\theta_{\emph{naive},j}=g\{E(Y\mid W^c_j=1)\}$ for $j=1,\ldots,J$, and $\btheta_{\emph{naive}}$ collects these category-specific parameters. Throughout, the true and observed categories under consideration have positive probability, and their corresponding category-specific means lie in the interior of the mean domain of $g$. Because measurement error can move individuals across the category cutpoints, the naive analysis targets $\btheta_{\emph{naive}}$, which generally differs from the error-free target $\btheta$ \citep{achic2018categorizing}.

Our primary estimand is the category contrast $\theta_J-\theta_1$, which compares the $J$th and first exposure categories. For a binary response with a logistic link, this contrast is the corresponding log odds ratio. Replicate exposure measurements provide information about the measurement-error mechanism needed to recover $\btheta$ from the response and error-prone exposure measurements. Section~\ref{subsec:simfex} develops the mean-scale relation underlying the SIMFEX correction, and Section~\ref{subsec:correlated} extends the construction to covariate-adjusted analyses. The extension to two error-prone continuous covariates is described in Appendix~2.3.

\section{Method}\label{sec:methodology}
\subsection{The SIMFEX method}\label{subsec:simfex}
SIMFEX estimates the category-specific parameters of the categorical misspecified model through a deterministic construction on the mean scale. Its key population identity expresses the observed-category means as a transformation of the error-free category means. Additional admissible powers of the misclassification matrix then generate a pseudo-mean trajectory that can be extrapolated to the no-misclassification endpoint. The mean-scale formulation provides a common correction for one-to-one links in the baseline setting.

We characterize the measurement error mechanism from $\bX^c$ to $\bW^c$ by the misclassification matrix $\Pi=(\pi_{j'j})_{J\times J}$, where
\begin{align}\label{eq:def_pi}
\pi_{j'j}
&=P(W^c_j=1\mid X^c_{j'}=1)
=P(W\in C_j\mid X\in C_{j'}).
\end{align}
The probability vector $\bp=(p_1,\ldots,p_J)^\top$ describes the true category distribution, with
\begin{equation}\label{eq:def_p}
p_{j'}=P(X^c_{j'}=1)=P(X\in C_{j'}).
\end{equation}
The standing positivity condition for the observed categories is equivalent to requiring that $\Pi^\top\bp$ have strictly positive components.

\begin{Assum}[Category-level mean nondifferentiality]\label{assum:catmean}
For every category pair $(j,j')$ with $P(X_{j'}^c=1,W^c_j=1)>0$,
\begin{equation*}
E(Y\mid X_{j'}^c=1,W^c_j=1)=E(Y\mid X_{j'}^c=1).
\end{equation*}
\end{Assum}
This condition concerns the category-specific conditional mean. For nonbinary outcomes, it is weaker than full conditional independence of $Y$ and $\bW^c$ given $\bX^c$; for binary outcomes, the two conditions are equivalent.

Define the mean-scale parameters
\[
\bmu=g^{-1}(\btheta),
\qquad
\bmu_{\emph{naive}}=g^{-1}(\btheta_{\emph{naive}}),
\]
where $g^{-1}$ is applied componentwise. Under Assumption~\ref{assum:catmean},
\begin{align}\label{eq:model_xw}
E(Y\mid W^c_j=1)
&=\sum_{j'=1}^J
g^{-1}(\theta_{j'})
P(X^c_{j'}=1\mid W^c_j=1),
\end{align}
where
\begin{align*}
P(X^c_{j'}=1\mid W^c_j=1)
=
\frac{\pi_{j'j}p_{j'}}{(\Pi_{\cdot j})^\top\bp},\qquad
\Pi_{\cdot j}
=
(\pi_{1j},\ldots,\pi_{Jj})^\top.
\end{align*}
The naive misspecified model similarly gives
\begin{equation}\label{eq:model_w}
E(Y\mid W^c_j=1)=g^{-1}(\theta_{\emph{naive},j}).
\end{equation}

For a misclassification matrix $Q$ and a probability vector $\bv$ such that $Q^\top\bv$ has strictly positive components, define
\begin{align}\label{eq:operator_A}
\cA(Q,\bv)
=
\{\operatorname{diag}(Q^\top\bv)\}^{-1}
Q^\top\operatorname{diag}(\bv).
\end{align}

\begin{Prop}[Mean-scale mixing relation]
Under Assumption~\ref{assum:catmean},
\begin{equation}\label{eq:theta_naive}
\bmu_{\textit{naive}}=\cA(\Pi,\bp)\bmu.
\end{equation}
When there is no measurement error, $\Pi=I$ and the relation reduces to $\bmu_{\textit{naive}}=\bmu$.
\end{Prop}

The link function does not enter the mixing operator \(\cA(\Pi,\bp)\) in eq~\eqref{eq:theta_naive}. SIMFEX therefore performs the correction on the mean scale and applies $g$ componentwise to obtain the corresponding coefficient-scale parameters.

Let $\mathcal E\subset[0,\infty)$ be an admissible extrapolation grid. For integer $\eta$, $\Pi^\eta$ is the usual matrix power. Following the matrix-power construction used in MCSIMEX \citep{kuchenhoff2006general}, for noninteger $\eta$ we define \(\Pi^\eta\) as the \(\eta\)th power of \(\Pi\) in the sense of spectral decomposition. We retain only grid points for which $\Pi^\eta$ is real, entrywise nonnegative, and row-stochastic, the corresponding $\cA$-operators are well defined, and 
\[
\Pi\Pi^\eta=\Pi^\eta\Pi=\Pi^{1+\eta}.
\]
At each $\eta\in\mathcal E$, define the pseudo-mean vector
\begin{equation}\label{eq:theta_naive_eta}
\bmu_{\emph{naive}}(\eta)
:=
\cA(\Pi^\eta,\Pi^\top\bp)\bmu_{\emph{naive}},
\end{equation}
where $\Pi^\top\bp$ is the observed category probability vector.

\begin{lemma}[Pseudo-mean trajectory]\label{lemma1}
For every $\eta\in\mathcal E$,
\begin{equation}\label{eq:A_composition}
\cA(\Pi^\eta,\Pi^\top\bp)\cA(\Pi,\bp)
=
\cA(\Pi^{1+\eta},\bp).
\end{equation}
Consequently, under Assumption~\ref{assum:catmean},
\begin{equation}\label{eq:formal_continuation_eta}
\bmu_{\textit{naive}}(\eta)
=
\cA(\Pi^\eta,\Pi^\top\bp)\bmu_{\textit{naive}}
=
\cA(\Pi^{1+\eta},\bp)\bmu.
\end{equation}
\end{lemma}

The composition in Lemma~\ref{lemma1} is summarized by the following flowchart:
\[
\begin{tikzcd}[column sep=4.1em]
\bmu
\arrow{r}{\scriptstyle\cA(\Pi,\bp)}
\arrow[bend right=22,dashed]{rr}{\scriptstyle\cA(\Pi^{1+\eta},\bp)}
&
\bmu_{\emph{naive}}
\arrow{r}{\scriptstyle\cA(\Pi^\eta,\Pi^\top\bp)}
&
\bmu_{\emph{naive}}(\eta).
\end{tikzcd}
\]
The two paths show that composing the original mixing with the additional $\Pi^\eta$ step replaces $\Pi$ by $\Pi^{1+\eta}$. The formal no-misclassification endpoint therefore corresponds to $\eta=-1$, for which $\Pi^{1+\eta}=\Pi^0=I$:
\[
\left.\cA(\Pi^{1+\eta},\bp)\bmu\right|_{\eta=-1}
=
\cA(I,\bp)\bmu
=
\bmu.
\]
SIMFEX estimates this endpoint by extrapolating from admissible nonnegative grid points. It does not require $\Pi^{-1}$ to be a valid misclassification matrix.

Let $\bh(\eta;\gamma)=\{h_1(\eta;\gamma),\ldots,h_J(\eta;\gamma)\}^\top$ denote a componentwise extrapolation function for $\bmu_{\emph{naive}}(\eta)$. Fitting $\bh$ at the selected grid points and evaluating it at $\eta=-1$ yields the SIMFEX estimator of $\bmu$; Algorithm~\ref{algo:simfex} gives the sample procedure.

\begin{Rem}[Extrapolation]\label{rem:extrapolate}
Common choices include linear, quadratic, and exponential functions. We use the quadratic form
\[
h_j(\eta;\gamma)
=
\gamma_{j1}+\gamma_{j2}\eta+\gamma_{j3}\eta^2,
\qquad
j=1,\ldots,J.
\]
Figure~\ref{fig:extra} illustrates the extrapolation step.
\end{Rem}

\begin{figure}
\centering
\includegraphics[scale=0.55]{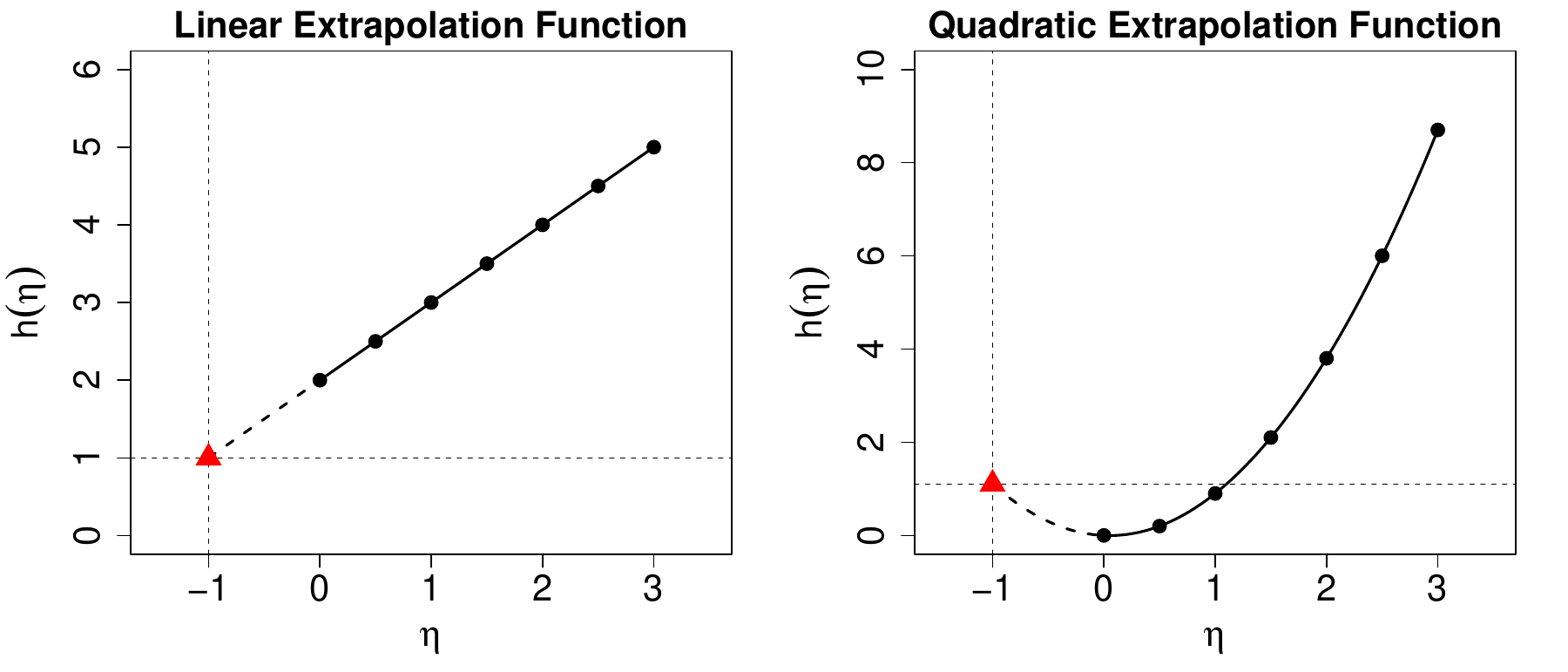}
\caption{Illustrations of linear and quadratic extrapolation functions. The black points mark function values at nonnegative values of $\eta$, the solid curves show the functions over this range, and the dashed curves extend them to $\eta=-1$. The red triangles mark the extrapolated values. These schematic points do not represent the grid used in the numerical studies.}
\label{fig:extra}
\end{figure}

\begin{Rem}[A bounded-perturbation characterization]
\label{rem:remainder}
The mean-scale construction also gives a quantitative characterization of departures from Assumption~\ref{assum:catmean}. By the law of total expectation,
\[
\bmu_{\emph{naive}}
=
\cA(\Pi,\bp)\bmu+\boldsymbol r,
\]
where
\begin{align*}
r_j
&=
\sum_{j'=1}^J
\cA(\Pi,\bp)_{jj'}\delta_{j'j}.
\end{align*}
For cells with $P(X^c_{j'}=1,W^c_j=1)>0$, define
\[
\delta_{j'j} = E(Y\mid X^c_{j'}=1,W^c_j=1) - E(Y\mid X^c_{j'}=1).
\]
Set $\delta_{j'j}=0$ for the remaining cells, whose weights in $r_j$ are zero.
Conditional independence of $Y$ and $W$ given $X$ does not by itself imply the category-level condition, because $W^c$ can retain information about the location of $X$ within a true category. Because each row of $\cA(\Pi,\bp)$ is a probability vector, $\max_{j',j}|\delta_{j'j}|\leq\Delta$ implies $\|\boldsymbol r\|_\infty\leq\Delta$. Thus, uniformly bounded within-category mean differences perturb the exact relation by at most $\Delta$ in supremum norm.
\end{Rem}

\begin{Rem}[Direct plug-in inversion]\label{rem:A_Pi}
The mean-scale mixing relation also yields a direct algebraic benchmark. When $\Pi$ is nonsingular and all components of $\bp$ are positive, $\cA(\Pi,\bp)$ is nonsingular and
\[
\cA(\Pi,\bp)^{-1}
=
\{\operatorname{diag}(\bp)\}^{-1}
(\Pi^\top)^{-1}
\operatorname{diag}(\Pi^\top\bp).
\]
Using the sample estimators introduced in Section~\ref{sec:algorithm}, define the direct-inversion estimator
\[
\widehat{\bmu}_{\mathrm{DI}}
=
\cA(\widehat\Pi,\widehat\bp)^{-1}
\widehat{\bmu}_{\emph{naive}}.
\]
When every component of $\widehat{\bmu}_{\mathrm{DI}}$ lies in the mean domain of $g$, define the corresponding coefficient-scale estimator by
\[
\widehat{\btheta}_{\mathrm{DI}}
=
g(\widehat{\bmu}_{\mathrm{DI}}).
\]
Here, $g$ is applied componentwise. If $\widehat\Pi\stackrel{p}{\to}\Pi$, $\widehat\bp\stackrel{p}{\to}\bp$, and $\widehat{\bmu}_{\emph{naive}}\stackrel{p}{\to}\bmu_{\emph{naive}}$, continuity of matrix inversion gives $\widehat{\bmu}_{\mathrm{DI}}\stackrel{p}{\to}\bmu$ whenever $\cA(\Pi,\bp)$ is nonsingular. Because $\bmu$ lies in the interior of the mean domain, the coefficient-scale estimator is then defined with probability tending to one, and continuity of $g$ at $\bmu$ gives its consistency. In finite samples, however, an ill-conditioned $\cA(\Pi,\bp)$ can amplify estimation error, and for bounded mean domains the inverse estimate can fall outside the domain of $g$. Appendix~3.4 provides a
numerical illustration of this instability. 

\end{Rem}

\subsection{Estimation and inference}\label{sec:algorithm}
SIMFEX estimation has three components. We first estimate the misclassification matrix $\Pi$ and the true category probability vector $\bp$ from replicate exposure information. We then evaluate the pseudo-mean trajectory deterministically and extrapolate it to $\eta=-1$ to estimate $\btheta$ and the category contrast $\theta_J-\theta_1$. Uncertainty is quantified by rerunning the complete procedure in subject-level bootstrap samples.

A distinctive feature of SIMFEX is that it separates estimation of the continuous measurement-error mechanism from correction of the categorized outcome model. We implement nuisance estimation with a Gaussian specification on the Box-Cox transformed scale \citep{wei2009quantile}:
\begin{align}\label{eq:mem_model}
\Lambda(W,\lambda)
=
\Lambda(X,\lambda)+U,\qquad
\Lambda(X,\lambda)
\sim N(\mu_{\lambda x},\sigma^2_{\lambda x}),
\qquad
U\sim N(0,\sigma^2_u).
\end{align}
Here, $U$ is independent of $X$.
For positive $x$, the Box-Cox transformation is
\[
\Lambda(x,\lambda)
=
\begin{cases}
(x^\lambda-1)/\lambda, & \lambda\neq0,\\
\log(x), & \lambda=0.
\end{cases}
\]
The choices $\lambda=1$ and $\lambda=0$ give the additive model $W=X+U$ and the multiplicative model $\log(W)=\log(X)+U$, respectively. The present specification is used for positive exposures, including the dietary exposure considered in Section~\ref{sec:data analysis}.
For $\lambda\neq0$, we use the untruncated Gaussian law in eq~\eqref{eq:mem_model} as a working specification on the transformed scale. In the simulation settings considered here, all generated transformed values remained in the range of the inverse Box-Cox transformation.

Let $\phi$ denote the standard normal density. The nuisance calculation uses
\begin{align}\label{eq:f_xw_x}
f(w\mid x)
&=
\sigma_u^{-1}
\phi\left\{
\frac{\Lambda(w,\lambda)-\Lambda(x,\lambda)}{\sigma_u}
\right\}
w^{\lambda-1},\\
f(x)
&=
\sigma_{\lambda x}^{-1}
\phi\left\{
\frac{\Lambda(x,\lambda)-\mu_{\lambda x}}{\sigma_{\lambda x}}
\right\}
x^{\lambda-1}.\notag
\end{align}
These expressions yield
\begin{align}
\pi_{j'j}
&=
\frac{
\int_{x\in C_{j'}}
\int_{w\in C_j}
f(w\mid x)f(x)\mathrm{d}w\mathrm{d}x
}{
\int_{x\in C_{j'}}f(x)\mathrm{d}x
},
\label{eq:pi}\\
p_j
&=
\int_{x\in C_j}f(x)\mathrm{d}x,
\qquad
j,j'=1,\ldots,J.
\label{eq:p}
\end{align}
From the fitted Box-Cox nuisance model, we estimate the category-level quantities $\Pi$ and $\bp$ that enter the SIMFEX correction. The outcome analysis categorizes the original-scale measurement $W$ as $\bW^c$ and fits the naive misspecified model in eq~\eqref{eq:misspecified_model_error} to estimate $\btheta_{\emph{naive}}$; SIMFEX then estimates the error-free target $\btheta$.

\begin{Rem}[Alternative nuisance models]\label{rem:hetergeneous}
SIMFEX uses the nuisance model through $\Pi$ and $\bp$. Alternative consistent estimators of these quantities can therefore be incorporated without rederiving the outcome correction. Across the known one-to-one links considered in the baseline setting, the pseudo-mean and extrapolation steps retain the same form. When the measurement error is heterogeneous, the nuisance estimator must account for that heterogeneity. When $\Pi$ and $\bp$ vary with error-free covariates, pooled and covariate-specific corrections have distinct roles, as discussed in Section~\ref{subsec:correlated}.
\end{Rem}

\begin{Assum}[Replicate information]\label{assum1}
Suppose that $R\geq2$ replicate exposure measurements are available either for subjects in the primary study or from an external replicate study. When an external study is used, the nuisance parameters $(\lambda,\mu_{\lambda x},\sigma^2_{\lambda x},\sigma^2_u)$ under model~\eqref{eq:mem_model} are common to the primary and external populations, so that the replicate study identifies $\Pi$ and $\bp$ for the primary study \citep[Chapter~2.2]{carroll2006nonlinear}.
\end{Assum} 

Under model~\eqref{eq:mem_model}, replicate measurements identify $\sigma^2_u$ by separating variation across repeated measurements from variation across subjects \citep{carroll2006nonlinear,wei2009quantile,yi2017statistical}. For presentation, we use the external-replicate notation. Let $(Y_i,W_i)$ denote the observations from the primary study, $i=1,\ldots,n$, and let $W_{i'r}$ denote the $r$th replicate for subject $i'$ in the external study, $i'=1,\ldots,n_0$ and $r=1,\ldots,R$. When repeated measurements are available in the primary study, the same nuisance-estimation steps are applied to those measurements.

We estimate the Box-Cox parameter $\lambda$ from the primary measurements by maximizing the profile loglikelihood \citep{box1964analysis}. Define
\begin{align*}
\overline\Lambda(\cdot,\lambda)
=n^{-1}\sum_{i=1}^n\Lambda(W_i,\lambda),\qquad
s_n^2(\lambda)
=n^{-1}\sum_{i=1}^n
\{\Lambda(W_i,\lambda)-\overline\Lambda(\cdot,\lambda)\}^2.
\end{align*}
Up to an additive constant,
\begin{align}\label{eq:lambda_hat}
\widehat{\lambda}
=\arg\max_{\lambda}L_n(\lambda),\qquad
L_n(\lambda)
=-\frac{n}{2}\log\{s_n^2(\lambda)\}
+(\lambda-1)\sum_{i=1}^n\log(W_i).
\end{align}
For external subject $i'$, define
\[
\overline\Lambda_{i'}
=R^{-1}\sum_{r=1}^R\Lambda(W_{i'r},\widehat\lambda)
\]
and let $\widehat\sigma^2_{u,i'}$ be the sample variance of the transformed replicates $\{\Lambda(W_{i'r},\widehat\lambda):r=1,\ldots,R\}$. The estimating equations for $(\mu_{\lambda x},\sigma^2_u,\sigma^2_{\lambda x})$ are
\begin{align}\label{eq:equations_sigma_u}
0
&=n_0^{-1}\sum_{i'=1}^{n_0}
(\overline\Lambda_{i'}-\mu_{\lambda x}),\\
0
&=n_0^{-1}\sum_{i'=1}^{n_0}
(\widehat\sigma^2_{u,i'}-\sigma^2_u),\notag\\
0
&=n_0^{-1}\sum_{i'=1}^{n_0}
\{(\overline\Lambda_{i'}-\mu_{\lambda x})^2
-\sigma^2_{\lambda x}-\sigma^2_u/R\}.\notag
\end{align}
The fitted exposure and error distributions then yield $\widehat\Pi$ and $\widehat\bp$ through eqs~\eqref{eq:f_xw_x}-\eqref{eq:p}. Algorithm~\ref{algo:pi_p} summarizes this nuisance-estimation step.

\begin{algorithm} 
\caption{Estimate the misclassification matrix $\Pi$ and the probability vector $\bp$} 
\label{algo:pi_p} 
\begin{algorithmic}
\State Given the primary measurements $(W_i)_{i=1}^n$, replicate measurements $\{(W_{i'r})_{r=1}^{R}\}_{i'=1}^{n_0}$ from the primary or an external study, and the prespecified categories $(C_1,\ldots,C_J)$.
\State \textbf{Step 1:} Set $\lambda$ to its prespecified value or estimate it via eq~\eqref{eq:lambda_hat};
\State \textbf{Step 2:} Estimate $\mu_{\lambda x}$, $\sigma^2_u$, and $\sigma^2_{\lambda x}$ via eq~\eqref{eq:equations_sigma_u};
\State \textbf{Step 3:} Evaluate the fitted densities $f(w\mid x)$ and $f(x)$ in eq~\eqref{eq:f_xw_x};
\State \textbf{Step 4:} Compute $\widehat\Pi$ and $\widehat\bp$ by numerical integration in eqs~\eqref{eq:pi}-\eqref{eq:p}.
\end{algorithmic}
\end{algorithm} 

\begin{Rem}[Nuisance estimation]
Algorithm~\ref{algo:pi_p} converts the fitted exposure and error distributions directly into $\widehat\Pi$ and $\widehat\bp$. If the nuisance parameters in model~\eqref{eq:mem_model} are consistently estimated, continuity of the integral mappings in eqs~\eqref{eq:pi}-\eqref{eq:p} yields $\widehat\Pi\stackrel{p}{\to}\Pi$ and $\widehat\bp\stackrel{p}{\to}\bp$. When the additive or multiplicative error form is specified, $\lambda$ is fixed at $1$ or $0$, respectively, and Step~1 is omitted. When $\lambda$ is estimated, Section~\ref{sec:simulation} evaluates the finite-sample performance of the resulting plug-in nuisance estimators.
\end{Rem}

With $\widehat\Pi$ and $\widehat\bp$ from Algorithm~\ref{algo:pi_p}, the remaining SIMFEX calculation has the same form across the known one-to-one links considered in the baseline setting. Given the primary data $(Y_i,W_i)_{i=1}^n$, we categorize $W_i$ as $\bW_i^c$ and fit the \emph{naive} misspecified model in eq~\eqref{eq:misspecified_model_error} to obtain $\widehat\btheta_{\emph{naive}}$ and set $\widehat\bmu_{\emph{naive}}=g^{-1}(\widehat\btheta_{\emph{naive}})$, with $g^{-1}$ applied componentwise.

For each admissible grid point $\eta_k$, $k=1,\ldots,K$, compute the pseudo-mean vector
\begin{equation}\label{eq:theta_eta_k}
\widehat\bmu_{\emph{naive}}(\eta_k)
=
\cA\left(
(\widehat\Pi)^{\eta_k},
\widehat\Pi^\top\widehat\bp
\right)
\widehat\bmu_{\emph{naive}}.
\end{equation}
This step evaluates the trajectory without simulating pseudo-datasets or refitting the outcome model. In the reported analyses, we use the grid $\{0.5,1,1.5,2\}$.

We fit the componentwise quadratic extrapolation described in Remark~\ref{rem:extrapolate} to $\{(\eta_k,\widehat\mu_{\emph{naive},j}(\eta_k)):k=1,\ldots,K\}$ for each $j=1,\ldots,J$. Let $\widehat\bgamma_j$ denote the fitted three-dimensional coefficient vector for category $j$, and let $\widehat\gamma=(\widehat\bgamma_1^\top,\ldots,\widehat\bgamma_J^\top)^\top$. The SIMFEX estimator on the mean scale is
\[
\widehat\bmu_{\emph{simfex}}
:=
\bh(-1;\widehat\gamma).
\]
Let $\mathcal D_g$ denote the mean domain of $g$, and fix any $\btheta_\star$ in the coefficient parameter space. Define the coefficient-scale estimator by applying $g$ componentwise whenever the extrapolated mean lies in its domain:
\begin{equation}\label{eq:theta_simfex}
\widehat\btheta_{\emph{simfex}}
:=
\begin{cases}
g(\widehat\bmu_{\emph{simfex}}),
& \widehat\bmu_{\emph{simfex}}\in\mathcal D_g^J,\\
\btheta_\star,
& \widehat\bmu_{\emph{simfex}}\notin\mathcal D_g^J.
\end{cases}
\end{equation}
Theorem~\ref{thm:consistent} shows that the fixed complementary-event convention is asymptotically immaterial. The estimator of the category contrast $\theta_J-\theta_1$ is $\widehat\theta_{\emph{simfex},J}-\widehat\theta_{\emph{simfex},1}$. Algorithm~\ref{algo:simfex} summarizes the complete point-estimation procedure.

\begin{algorithm}[!ht]
\caption{Point estimation with SIMFEX} 
\label{algo:simfex}
\begin{algorithmic}
\State Given the primary data $(Y_i,W_i)_{i=1}^n$, replicate exposure information in Assumption~\ref{assum1}, and prespecified categories $(C_1,\ldots,C_J)$.
\State \textbf{Step 1:} Estimate $\widehat\Pi$ and $\widehat\bp$ via Algorithm~\ref{algo:pi_p};
\State \textbf{Step 2:} Fit the \emph{naive} misspecified model and compute $\widehat\bmu_{\emph{naive}}(\eta_k)$ at each admissible grid point via eq~\eqref{eq:theta_eta_k};
\State \textbf{Step 3:} Fit the componentwise extrapolation functions and evaluate them at $\eta=-1$ to obtain $\widehat\bmu_{\emph{simfex}}$;
\State \textbf{Step 4:} Compute $\widehat\btheta_{\emph{simfex}}$ via eq~\eqref{eq:theta_simfex} and report the desired category contrasts.
\end{algorithmic}
\end{algorithm}

\begin{theorem}[Consistency of SIMFEX]\label{thm:consistent}
As $n\to\infty$ and, when an external replicate study is used, $n_0\to\infty$, suppose Assumptions~\ref{assum:catmean} and~\ref{assum1} hold. Further assume that:
\begin{enumerate}
    \item[(i)] The required estimators are consistent:
    \[
        \widehat\Pi\stackrel{p}{\to}\Pi,\qquad
        \widehat\bp\stackrel{p}{\to}\bp,\qquad
        \widehat\bmu_{\textit{naive}}\stackrel{p}{\to}\bmu_{\textit{naive}}.
    \]
    With probability tending to one, $\widehat\Pi$ is row stochastic and $\widehat\bp$ lies in the probability simplex.
    \item[(ii)] Fix a finite candidate grid \(\mathcal E=\{\eta_1,\ldots,\eta_K\}\subset[0,\infty)\), with $K\geq3$. For each $k$, there is a relative neighborhood \(\mathcal N_k\) of \((\Pi,\bp)\) within the product of the row-stochastic matrix space and the probability simplex such that, for every \((Q,\bv)\in\mathcal N_k\), the power \(Q^{\eta_k}\) is real-valued, continuous in $Q$, and an admissible misclassification matrix, and \(\cA(Q^{\eta_k},Q^\top\bv)\) is well defined. The quadratic design matrix with $k$th row $(1,\eta_k,\eta_k^2)$ has full column rank.
    \item[(iii)] The extrapolation family represents the population trajectory on the grid and at the extrapolation point: there exists $\gamma_0=(\bgamma_{01}^\top,\ldots,\bgamma_{0J}^\top)^\top$ such that for $k=1,\ldots,K$,
    \[
        \bmu_{\textit{naive}}(\eta_k)=\bh(\eta_k;\gamma_0),\qquad
        \bh(-1;\gamma_0)=\bmu .
    \]
    \item[(iv)] Let $\mathcal D_g$ denote the mean domain of $g$. The vector $\bmu$ lies in the interior of $\mathcal D_g^J$, and $g$ is continuous at $\bmu$.
\end{enumerate}
On samples for which the retained grid does not yield the full-rank quadratic fit, define $\widehat\bmu_{\textit{simfex}}$ to be any fixed vector. Conditions~(i) and~(ii) imply that this exceptional event has probability tending to zero.
Then
\[
    \widehat\bmu_{\textit{simfex}}\stackrel{p}{\to}\bmu .
\]
Moreover,
\[
P(\widehat\bmu_{\textit{simfex}}\in\mathcal D_g^J)\to1.
\]
Consequently, the coefficient-scale estimator in eq~\eqref{eq:theta_simfex} satisfies
\[
    \widehat\btheta_{\textit{simfex}}
    \stackrel{p}{\to}
\btheta.
\]
\end{theorem}
Proofs of Lemma~\ref{lemma1} and Theorem~\ref{thm:consistent} are given in Appendix~2.2. 

\begin{Rem}[Relation to SIMEX and MCSIMEX]
Theorem~\ref{thm:consistent} establishes consistency for the parameters of the categorical misspecified model through the deterministic SIMFEX construction. By representing the category-level distortion through $\Pi$ and $\bp$, the result provides asymptotic justification for deterministic correction without pseudo-dataset simulation, complementing the asymptotic theory for SIMEX and the extrapolation construction of MCSIMEX \citep{stefanski1995simulation,kuchenhoff2006general}. Condition~(iii) specifies when the chosen extrapolation family represents the population trajectory on the grid and at $\eta=-1$. Section~\ref{sec:simulation} evaluates the finite-sample performance of the quadratic extrapolation used in practice.
\end{Rem}

We use a subject-level bootstrap for uncertainty quantification. When the primary study contains repeated measurements, subjects are resampled together with their outcomes and all replicate measurements. When the primary and external replicate studies are separate, subjects are resampled independently within each study, with all replicates retained for each sampled subject. In every bootstrap sample, we re-estimate the Box-Cox parameter when it is not fixed, the measurement error variance components, $\Pi$, and $\bp$, refit the \emph{naive} misspecified model, and repeat the deterministic extrapolation.

The empirical standard deviations of the resulting SIMFEX estimates provide the standard errors, and 95\% confidence intervals are constructed by the normal approximation. This resampling calculation propagates uncertainty through all estimation stages. Related SIMEX and MCSIMEX variance calculations use analytic, jackknife, or asymptotic constructions \citep{stefanski1995simulation,kuchenhoff2006general,kuchenhoff2007asymptotic}; the bootstrap provides a common uncertainty-quantification procedure for SIMFEX.

\subsection{The SIMFEX method with error-free covariates}\label{subsec:correlated} 
SIMFEX retains an exact covariate-adjusted formulation for identity-link additive mean models. When the misclassification mechanism and the true category probabilities are invariant with respect to the error-free covariates, the covariate contribution is common across the observed exposure categories and cancels from category contrasts. The resulting exact population contrast relation supports the same deterministic extrapolation construction used in Section~\ref{subsec:simfex}. After establishing this result, we distinguish two separate approximation boundaries: a nonidentity link with covariates, and the use of pooled nuisance quantities when the nuisance structure varies with the covariates.

For almost every $\bz$ in the support of $\bZ$, define the conditional misclassification matrix from $\bX^c$ to $\bW^c$ by $\Pi(\bz)=\big(\pi_{j'j}(\bz)\big)_{J\times J}$, where
\[
\pi_{j'j}(\bz)=P(W^c_j=1\mid X^c_{j'}=1,\bZ=\bz).
\]
Also define $\bp(\bz)=(p_1(\bz),\ldots,p_J(\bz))^\top$, where, for $j=1,\ldots,J$,
\[
p_j(\bz)=P(X^c_j=1\mid\bZ=\bz)=P(X\in C_j\mid\bZ=\bz).
\]

We assume the conditional version of the category-level identifying condition given in Assumption~\ref{assum:catmean}.
\begin{Assum}[Conditional category-level mean nondifferentiality]
\label{assum:catmean_z}
For all category pairs \((j,j')\) and for almost every $\bz$ such that $P(X^c_{j'}=W^c_j=1\mid \bZ=\bz)>0$,
\[
    E(Y\mid X^c_{j'}=W^c_j=1,\bZ=\bz)
    =
    E(Y\mid X^c_{j'}=1,\bZ=\bz).
\]
\end{Assum}

Throughout this subsection, $\btheta$ denotes the covariate-adjusted category coefficient, which generally differs from the marginal target without $\bZ$ in Section~\ref{subsec:notationmodel}. The vector $\bZ$ does not include an intercept because all $J$ category indicators are retained in the model.

Under the identity link, define the covariate-adjusted target parameter $\btheta$ through
\begin{equation}\label{eq:misspecifiedmodel_z}
E(Y\mid \bX^c,\bZ)
=
\btheta^\top\bX^c+\btheta_z^\top\bZ .
\end{equation}
The \emph{naive} method now substitutes $\bW^c$ for $\bX^c$ directly and models 
\begin{equation}\label{eq:misspecified_model_error_z}
    E(Y\mid \bW^c,{\bZ}) = \btheta_{\emph{naive}}^\top{\bW^c} + \btheta_{z,\emph{naive}}^\top\bZ.
\end{equation}
For the exact global coefficient-level relation, suppose that the conditional nuisance structure is invariant in $\bZ$, so that $\Pi(\bz)\equiv\Pi$ and $\bp(\bz)\equiv\bp$, with $p_j>0$ for $j=1,\ldots,J$. One sufficient condition is that $X$ is independent of $\bZ$ and that the measurement error mechanism does not depend on $\bZ$ conditional on $X$. Under this invariance, the posterior category weights given $\bW^c$ and $\bZ$ do not depend on $\bz$.

Based on eq~\eqref{eq:misspecifiedmodel_z}, under Assumption~\ref{assum:catmean_z}, for almost every $\bz$ and each $j=1,\ldots,J$,
\begin{align}\label{eq:model_xw-2}
E(Y\mid W^c_j=1,\bZ=\bz) =
\sum_{j'=1}^J
\theta_{j'}
P(X^c_{j'}=1\mid W^c_j=1,\bZ=\bz)
+
\btheta_z^\top\bz,
\end{align}
where
\begin{align*}
P(X^c_{j'}=1\mid W^c_j=1,\bZ=\bz)
=
\frac{\pi_{j'j}p_{j'}}
{\Pi_{\cdot j}^\top\bp}.
\end{align*}
Meanwhile, based on eq~\eqref{eq:misspecified_model_error_z},
\begin{equation}\label{eq:model_w-2}
E(Y\mid W^c_j=1,\bZ=\bz)
=
\theta_{\emph{naive},j}
+
\btheta^\top_{z,\emph{naive}}\bz.
\end{equation}
Under the invariance condition, the right-hand side of eq~\eqref{eq:model_xw-2} is of the naive model form. Its category coefficient vector is $\cA(\Pi,\bp)\btheta$, and its covariate coefficient is $\btheta_z$. For each $j=2,\ldots,J$, subtracting the equality for the reference category $j=1$ removes the common covariate contribution. Thus, eqs~\eqref{eq:model_xw-2}-\eqref{eq:model_w-2} yield
\begin{equation}\label{eq:relation_z}
\theta_{\emph{naive},j}-\theta_{\emph{naive},1}
=
\sum_{j'=1}^J
\left(
\frac{\pi_{j'j}p_{j'}}{\Pi_{\cdot j}^\top\bp}
-
\frac{\pi_{j'1}p_{j'}}{\Pi_{\cdot 1}^\top\bp}
\right)
\theta_{j'}.
\end{equation}
To express these $J-1$ relations compactly, define
\[
C=
\begin{pmatrix}
-\mathbf 1_{J-1} & I_{J-1}
\end{pmatrix},
\qquad
L=
\begin{pmatrix}
\mathbf 0_{J-1}^{\top}\\
I_{J-1}
\end{pmatrix}.
\]
For an admissible misclassification matrix $Q$ and probability vector $\bv$, let
\begin{equation}\label{eq:operator_tilde_A}
\widetilde\cA(Q,\bv)
=
C\cA(Q,\bv)L
\in\mathbb R^{(J-1)\times(J-1)}.
\end{equation}
Also define
\begin{align*}
\widetilde\btheta
&:=C\btheta
=(\theta_2-\theta_1,\ldots,\theta_J-\theta_1)^\top,\\
\widetilde\btheta_{\emph{naive}}
&:=C\btheta_{\emph{naive}}
=(\theta_{\emph{naive},2}-\theta_{\emph{naive},1},
\ldots,
\theta_{\emph{naive},J}-\theta_{\emph{naive},1})^\top.
\end{align*}
Because $\btheta=\theta_1\mathbf 1_J+L\widetilde\btheta$ and every row of $\cA(\Pi,\bp)$ sums to one, the common category level is removed by $C$. Eq~\eqref{eq:relation_z} therefore gives the exact global population relation
\begin{equation}\label{eq:theta_naive-2}
\widetilde\btheta_{\emph{naive}}
=
\widetilde\cA(\Pi,\bp)\widetilde\btheta.
\end{equation}
Thus, after the common covariate contribution is removed, the target and naive category contrasts satisfy the exact matrix relation in eq~\eqref{eq:theta_naive-2}.

For each $\eta$ in the same admissible matrix-power grid used in Section~\ref{subsec:simfex}, define
\begin{equation}\label{eq:theta_naive_eta_z}
\widetilde\btheta_{\emph{naive}}(\eta)
:=
\widetilde\cA(\Pi^\eta,\Pi^\top\bp)
\widetilde\btheta_{\emph{naive}}.
\end{equation}
The row-stochastic property of the corresponding $\cA$-operators and Lemma~\ref{lemma1} yield the contrast-level composition identity
\begin{equation}\label{eq:Atilde_composition}
\widetilde\cA(\Pi^\eta,\Pi^\top\bp)
\widetilde\cA(\Pi,\bp)
=
\widetilde\cA(\Pi^{1+\eta},\bp).
\end{equation}
Consequently,
\[
\widetilde\btheta_{\emph{naive}}(\eta)
=
\widetilde\cA(\Pi^{1+\eta},\bp)\widetilde\btheta,
\]
whose formal value at $\eta=-1$ is $\widetilde\cA(I,\bp)\widetilde\btheta=\widetilde\btheta$. As in the baseline construction, $\eta=-1$ is an extrapolation point rather than a grid point at which $\Pi^{-1}$ is used directly.

For estimation, let
\[
\widehat{\widetilde{\btheta}}_{\emph{naive}}
=
C\widehat\btheta_{\emph{naive}}.
\]
For each admissible grid point $\eta_k$, compute
\begin{equation}\label{eq:theta_eta_k_z}
\widehat{\widetilde{\btheta}}_{\emph{naive}}(\eta_k)
=
\widetilde\cA\left((\widehat\Pi)^{\eta_k},(\widehat\Pi)^\top\widehat\bp\right)
\widehat{\widetilde{\btheta}}_{\emph{naive}}.
\end{equation}
We fit $J-1$ componentwise extrapolation functions to these contrast trajectories and evaluate them at $\eta=-1$, producing $\widehat{\widetilde{\btheta}}_{\emph{simfex}}$. Its components estimate $\theta_j-\theta_1$ for $j=2,\ldots,J$. Under the stated invariance condition, eq~\eqref{eq:theta_naive-2} is an exact population contrast relation, and its formal no-misclassification endpoint is $\widetilde\btheta$. The sample estimator replaces the nuisance quantities and the naive contrast by their estimates and uses the selected extrapolation family to estimate that endpoint.

For a nonidentity link with a covariate-invariant nuisance structure, SIMFEX applies the deterministic matrix extrapolation after the nuisance quantities are estimated. In the simulations and the UK Biobank analysis, a single fit of the covariate-adjusted \emph{naive} generalized linear model provides the fitted category coefficients. The extrapolation vector consists of their componentwise inverse-link transformations, with the fitted coefficients of $\bZ$ treated as nuisance coefficients rather than extrapolation targets.

For nonidentity links with error-free covariates, averaging over the latent exposure categories produces a covariate-specific nonlinear mixture. By the same conditioning argument used for eq~\eqref{eq:model_xw-2}, for almost every $\bz$,
\begin{align*}
    E(Y\mid W^c_j = 1,\bZ=\bz) = \sum^J_{j'=1}g^{-1}(\theta_{j'} + \btheta^\top_z\bz) P(X^c_{j'}=1\mid W^c_j=1,\bZ=\bz).
\end{align*}
Because $g^{-1}$ is nonlinear, this mixture generally falls outside the fitted \emph{naive} generalized linear model family
$
    g^{-1}(\theta_{\emph{naive},j}+\btheta_{z,\emph{naive}}^\top\bz).
$
Like SIMEX and MCSIMEX, SIMFEX addresses this setting through extrapolation, but does so deterministically, without generating pseudo-data or repeatedly refitting the outcome model \citep{cook1994simex,kuchenhoff2006general}. We use the resulting estimator as a practical approximation.

The pooled procedure estimates one misclassification matrix $\Pi$ and one category-probability vector $\bp$ and uses these quantities in the deterministic SIMFEX extrapolation. For the identity-link model, this procedure is exact under the invariance condition used above. When $\Pi(\bz)$ or $\bp(\bz)$ varies with $\bz$, replacing these conditional quantities by pooled values introduces a second working approximation, distinct from the nonidentity-link approximation. Under the measurement error model in eq~\eqref{eq:mem_model}, the conditional nuisance quantities are
\begin{align*}
\pi_{j'j}(\bz)
&=
\frac{\int_{x\in C_{j'}}\int_{w\in C_j}
f(w\mid x,\bz)f(x\mid\bz)\mathrm{d}w\mathrm{d}x}
{\int_{x\in C_{j'}}f(x\mid\bz)\mathrm{d}x},\\
p_j(\bz)
&=
\int_{x\in C_j}f(x\mid\bz)\mathrm{d}x,
\qquad j,j'=1,\ldots,J.
\end{align*}
If the measurement error mechanism does not depend on $\bZ$ conditional on $X$, then $f(w\mid x,\bz)=f(w\mid x)$, and the first expression simplifies accordingly. A covariate-specific correction requires estimating $f(x\mid\bz)$ and, when needed, $f(w\mid x,\bz)$, then carrying the resulting $\bz$-specific misclassification matrices and category probabilities through the correction. For continuous or multivariate $\bZ$, this replaces one matrix and one probability vector by conditional functions over the covariate support and requires sufficient replicate information across that support.

The numerical studies in Section~\ref{subsec:simu_additional} show that SIMFEX reduces absolute bias relative to the \emph{naive} analysis and yields empirical coverage near the nominal level in the nonidentity-link and pooled-nuisance settings examined. Section~\ref{sec:data analysis} reports diagnostics comparing pooled and covariate-specific nuisance estimates in the application.

\subsection{Connection with other methods}\label{subsec:connection}
\citet{achic2018categorizing} and SIMFEX address the same scientific problem: estimation of category-level associations when an underlying continuous exposure is measured with error and then categorized. The methods differ in how continuous-scale measurement error information enters the outcome correction. \citet{achic2018categorizing} begin with a continuous-scale outcome model and derive observed-data estimating equations involving conditional expectations of the latent exposure $X$, whose explicit forms depend on the selected outcome and measurement-error models. SIMFEX instead defines the target through category-specific means and, under Assumption~\ref{assum:catmean}, uses the mean-scale relation $\bmu_{\emph{naive}}=\cA(\Pi,\bp)\bmu$. In the baseline setting without additional error-free covariates, once $\Pi$ and $\bp$ are estimated, the same deterministic mean-scale transformation and extrapolation procedure applies for any known one-to-one link, followed by the componentwise transformation $\btheta=g(\bmu)$. This separation yields a common simulation-free correction without deriving a new observed-data estimating equation for each outcome model.

SIMFEX builds on the extrapolation principle shared by SIMEX and MCSIMEX for the present categorized-exposure problem. SIMEX adds simulated measurement error to the observed continuous covariate, repeatedly refits the estimator, and extrapolates the resulting trajectory to $\eta=-1$, the formal error-free point \citep{stefanski1995simulation}. After the exposure is categorized, the relevant distortion is the induced misclassification between $\bX^c$ and $\bW^c$, which calls for an extrapolation construction that directly targets the categorized predictor and its category-level parameters.

MCSIMEX provides the closest categorical analogue because it operates on a misclassified categorical covariate \citep{kuchenhoff2006general}. Given an available misclassification matrix $\Pi$, it uses selected powers $\Pi^\eta$ to simulate pseudo-categorized covariates, repeatedly refits the \emph{naive} model, and extrapolates the averaged estimator trajectory to $\eta=-1$. This adapts the SIMEX principle to categorical misclassification, while continuing to rely on Monte Carlo generation and repeated outcome-model fitting.

For the present categorized-exposure problem, Algorithm~\ref{algo:pi_p} estimates both $\Pi$ and $\bp$ from replicate error-prone measurements of the underlying continuous exposure. Given these nuisance quantities, SIMFEX evaluates the pseudo-mean trajectory deterministically through $\bmu_{\emph{naive}}(\eta)=\cA(\Pi^\eta,\Pi^\top\bp)\bmu_{\emph{naive}}$, replacing Monte Carlo generation and repeated outcome-model fitting with a direct calculation on the category-mean scale. Together, these two steps separate estimation of the measurement-error mechanism for the continuous exposure from downstream outcome correction.

\section{Simulation studies}\label{sec:simulation}
\subsection{Simulation settings}\label{subsec:simu_setting}

The simulation studies ask how well SIMFEX recovers a category contrast and quantifies its uncertainty when an underlying continuous exposure-response relationship is fitted using a categorized exposure. The primary $J=5$ study compares the point and interval performance of SIMFEX with two alternatives across several outcome models and exposure distributions. Supporting studies examine estimation of the nuisance quantities $\Pi$ and $\bp$, sensitivity to the magnitude of measurement error, and performance with additional error-free covariates.

In every design, the response is generated from a model involving the underlying continuous $X$, whereas the fitted model categorizes the exposure and estimates the parameters of the misspecified model in eq~\eqref{eq:misspecifiedmodel}. Thus, misspecification refers specifically to replacing the underlying continuous exposure-response relationship by a categorical model, a common empirical choice when category-specific effects are desired for interpretation. The primary study targets the error-free category contrast $\theta_5-\theta_1$.

It uses $J=5$ categories defined by fixed cutpoints that approximate the population quintiles of $X$. The cutpoints are computed once from a large reference draw and held fixed across Monte Carlo replicates. Each subject contributes two error-prone exposure replicates with independent measurement errors. The first replicate is categorized and used in the fitted outcome model, while the replicate pair provides the information used to estimate $\Pi$ and $\bp$. The latent $X$ is used to generate $Y$ and define the error-free target, but is not used by any of the fitted methods. The sample size is $n=1{,}000$, and the point and interval results are based on $N=500$ Monte Carlo replicates. We report Monte Carlo bias, root mean squared error (RMSE), average estimated standard error, and empirical coverage of nominal 95\% confidence intervals.

We compare SIMFEX with the \emph{naive} method and MCSIMEX*. The latter denotes MCSIMEX using the misclassification matrix estimated by Algorithm~\ref{algo:pi_p}. MCSIMEX* uses $B=100$ simulations at each grid point and the same quadratic extrapolation function and grid $\eta_k=0.5,1,1.5,2$ as SIMFEX. Thus, the point-estimation comparison uses the same estimated misclassification matrix and extrapolation specification, while MCSIMEX* retains pseudo-dataset simulation and repeated outcome-model fitting. Standard errors for the \emph{naive} and SIMFEX estimators are computed using 500 bootstrap resamples, whereas standard errors for MCSIMEX* are calculated using jackknife variance estimation \citep{kuchenhoff2006general}. The interval comparisons therefore also reflect the different variance-estimation procedures.

We examine performance under three Box-Cox settings chosen to mimic approximately Gaussian and right-skewed exposures, as well as exposures with heavier tails. In all three settings, $\Lambda(X,\lambda)$ is Gaussian on the transformed scale. The fixed values $\lambda=1,0.26,0.95$ define Settings I, II, and III, respectively. Appendix~3 describes the calibration of the latter two values. 

The observed contaminated exposure $W$ is then generated from the additive measurement error model on the transformed scale,
$
    \Lambda(W,\lambda) = \Lambda(X,\lambda) + U,\ U\sim N(0,\sigma^2_u), 
$
where $\sigma^2_u$ is calibrated to the specified noise-to-signal ratio (NSR) at 0.8 and 1. Here, \(\mathrm{NSR}=(\sigma_w^2-\sigma_x^2)/\sigma_x^2\), where
\(\sigma_x^2=\operatorname{Var}(X)\) and
\(\sigma_w^2=\operatorname{Var}(W)\) are evaluated on the original exposure scale rather than the Box-Cox transformed scale. In each Monte Carlo replicate, SIMFEX estimates $\lambda$ from the observed $W$ values using the \texttt{boxcox} function in the \texttt{R} package \texttt{MASS}. The remaining exposure and measurement error parameters are reported in Appendix~3. The response $Y$ is generated from three continuous-exposure models: 
\begin{enumerate}
    \item Logistic regression:
     $
        H^{-1}\{P(Y=1\mid X)\} = \beta_0 + \beta_1X,\quad H(t) = e^t/(1+e^t).
     $
    \item Linear regression:
    $
        Y = \beta_0 + \beta_1X + \epsilon,\quad \epsilon \sim N(0,0.75^2).
    $
    \item Probit regression:
    $
        \Phi^{-1}\{P(Y=1\mid X)\} = \beta_0 + \beta_1X,\quad \Phi(t) = (2\pi)^{-1/2}\int^t_{-\infty}e^{-u^2/2}\mathrm{d}u.
    $
\end{enumerate}

We use the same link in the data-generating and fitted models for simplicity, although the SIMFEX construction does not require them to coincide. The primary study uses ${\rm NSR}=0.8$. We also examine a simpler setting with ${\rm NSR}=1$ and $J=3$ categories defined by the tertiles of $X$; its complete design and results are reported in Appendix~3.3. Detailed values of $\beta_0$, $\beta_1$, and the target contrasts are given in Appendix~3. 

\subsection{Results}\label{subsec:simu_result}
Table~\ref{tab:snr0.8J5} summarizes point estimation of $\theta_5-\theta_1$ in the primary $J=5$ study. Both correction methods substantially reduce the absolute bias of the \emph{naive} estimator. Relative to MCSIMEX*, SIMFEX has smaller RMSE in all nine settings, while MCSIMEX* has slightly smaller absolute bias.

\begin{table}
\begin{center}
\caption{Monte Carlo bias and root mean squared error for $\theta_5-\theta_1$ with $J=5$ and ${\rm NSR}=0.8$.}\label{tab:snr0.8J5} 
\resizebox{\columnwidth}{!}{
\begin{tabular}{cccccccc}
\toprule
 & &  \multicolumn{3}{c}{\bfseries \normalsize Average Bias}  & 
    \multicolumn{3}{c}{\bfseries \normalsize RMSE}\vspace{-1mm}\\
    \cmidrule(r){3-5} \cmidrule(r){6-8} 
    Model & Setting & \emph{naive} & MCSIMEX* & SIMFEX & \emph{naive} & 
    MCSIMEX* & SIMFEX\\ 
    \midrule
    logistic
    & I   & -0.36 & -0.04 & -0.06 & 0.46 & 0.49 & 0.47 \\ 
    & II  & -0.21 & -0.02 & -0.05 & 0.29 & 0.31 & 0.26 \\ 
    & III & -0.21 & -0.04 & -0.07 & 0.29 & 0.33 & 0.28 \\ 
    \midrule
    linear
    & I   & -0.10 & -0.03 & -0.04 & 0.12 & 0.11 & 0.10 \\ 
    & II  & -0.18 & -0.02 & -0.05 & 0.19 & 0.12 & 0.10 \\ 
    & III & -0.13 & -0.03 & -0.05 & 0.15 & 0.11 & 0.10 \\ 
    \midrule
    probit
    & I   & -0.18 & -0.02 & -0.04 & 0.23 & 0.25 & 0.23 \\ 
    & II  & -0.09 & -0.01 & -0.02 & 0.15 & 0.20 & 0.17 \\ 
    & III & -0.09 & -0.01 & -0.02 & 0.14 & 0.19 & 0.15 \\ 
    \bottomrule
\end{tabular}
}
\end{center}
\end{table}

Table~\ref{tab:vesnr0.8J5} reports the average estimated standard errors and empirical coverage rates. SIMFEX coverage ranges from 91\% to 96\% and is closer to the nominal 95\% level than the coverage of either comparison method in all nine settings. In the linear settings, its coverage is 91\% to 92\%, below nominal but higher than the corresponding coverage of both comparison methods. The SIMFEX standard errors are generally larger than those from the \emph{naive} analysis, reflecting the additional uncertainty introduced by measurement error correction.

\begin{table}
\begin{center}
\caption{Average estimated standard errors and empirical coverage of nominal 95\% confidence intervals for $\theta_5-\theta_1$ with $J=5$ and ${\rm NSR}=0.8$.}\label{tab:vesnr0.8J5}
\resizebox{\columnwidth}{!}{
\begin{tabular}{cccccccc}
    \toprule
    & &  \multicolumn{3}{c}{\bfseries \normalsize Standard Error}  & 
    \multicolumn{3}{c}{\bfseries \normalsize Coverage Rate}\vspace{-1mm}\\
    \cmidrule(r){3-5} \cmidrule(r){6-8} 
    Model & Setting & \emph{naive} & MCSIMEX* & SIMFEX & \emph{naive} & 
    MCSIMEX* & SIMFEX\\ 
    \midrule
    logistic
    & I   & 0.28 & 0.39 & 0.48 & 0.73 & 0.88 & 0.96 \\
    & II  & 0.18 & 0.24 & 0.24 & 0.78 & 0.87 & 0.94 \\
    & III & 0.19 & 0.26 & 0.27 & 0.78 & 0.90 & 0.94 \\
    \midrule
    linear
    & I   & 0.07 & 0.09 & 0.09 & 0.68 & 0.87 & 0.92 \\ 
    & II  & 0.07 & 0.09 & 0.09 & 0.25 & 0.86 & 0.91 \\ 
    & III & 0.07 & 0.09 & 0.09 & 0.45 & 0.86 & 0.91 \\ 
    \midrule
    probit
    & I   & 0.15 & 0.20 & 0.23 & 0.76 & 0.89 & 0.96 \\ 
    & II  & 0.13 & 0.17 & 0.17 & 0.90 & 0.90 & 0.95 \\ 
    & III & 0.11 & 0.15 & 0.15 & 0.87 & 0.89 & 0.94 \\ 
    \bottomrule
\end{tabular}
}
\end{center}
\end{table}

Taken together, the primary study shows that SIMFEX combines substantial bias correction with a favorable finite-sample trade-off: it has smaller RMSE than MCSIMEX* and coverage closer to the nominal level in all nine settings. SIMFEX obtains its extrapolation trajectory through deterministic calculations, without pseudo-dataset simulation or repeated outcome-model fitting.

A separate nuisance-estimation study supports the first step of the procedure. Based on $N=1{,}000$ Monte Carlo replicates, the mean and standard deviation of the Frobenius norm  $\|\widehat{\Pi}-\Pi\|_F$ are at most 0.020 and 0.009, respectively, and the corresponding values for $\|\widehat\bp-\bp\|_2$ are at most 0.022 and 0.012. The true and estimated nuisance quantities are reported in Appendix~3. 

The simpler $J=3$ study shows the same broad pattern as the primary study: both correction methods substantially reduce the bias of the \emph{naive} estimator, and SIMFEX coverage ranges from 93\% to 98\%. Complete results are reported in Tables~A4 and A5 in Appendix~3.3.

\subsection{Sensitivity analysis}\label{subsec:simu_sensity}
The primary study fixes the measurement-error magnitude at ${\rm NSR}=0.8$. We next ask how the finite-sample gains from correction change as measurement error becomes weaker or stronger. The sensitivity analysis uses the logistic model under Setting~I with $J=5$ and considers ${\rm NSR}=1,0.8,0.5,0.2$. For each NSR value, the results are based on $N=500$ Monte Carlo replicates and include Monte Carlo bias, RMSE, average estimated standard error, and empirical coverage for the category parameters $\theta_j$, $j=1,\ldots,5$, and four selected contrasts.

SIMFEX provides its clearest bias reduction when measurement error is moderate or large, particularly for the extreme-category parameters and contrasts. Its coverage remains between 0.92 and 1.00 across all reported NSR values. Undercoverage of the \emph{naive} method is most pronounced for the extreme-category quantities at ${\rm NSR}=1$ and 0.8, while MCSIMEX* undercovers several coefficients and contrasts across all four NSR values.

As measurement error decreases, attenuation of the \emph{naive} estimator becomes weaker, leaving less bias to correct. At ${\rm NSR}=0.2$, SIMFEX modestly overcorrects some upper-category parameters and contrasts; the additional variability from correction also means that lower bias need not yield lower RMSE. Detailed results are reported in Tables~A7 and A8 in Appendix~4.

\subsection{Covariate-adjusted simulation studies}\label{subsec:simu_additional}
We next ask whether the main empirical pattern extends to models that include an error-free scalar covariate $Z$. In the first set of designs, $X$ and $Z$ are generated independently and the measurement-error mechanism does not depend on $Z$, so that $\Pi(\bz)\equiv\Pi$ and $\bp(\bz)\equiv\bp$. The linear analysis uses the exact identity-link relation, whereas the logistic analysis uses the nonidentity-link approximation. Both models use the three exposure settings under $({\rm NSR},J)=(0.8,5)$ and $(1,3)$.

SIMFEX substantially reduces the bias of the \emph{naive} estimator in both analyses. For $J=5$, it has smaller RMSE than MCSIMEX* in all six combinations of outcome model and exposure setting, and across the reported $J=5$ and $J=3$ designs its coverage ranges from 0.91 to 0.97. Complete results based on $N=500$ Monte Carlo replicates are reported in Tables~A10 and~A11 in Appendix~5.

We then consider an application-motivated design in which $X$ follows sex-specific distributions and is generated independently of age. The continuous and binary outcomes correspond to BMI and obesity, respectively. SIMFEX approximately halves the absolute bias of the \emph{naive} estimator for both outcomes, has smaller RMSE than MCSIMEX* in both analyses, and achieves coverage rates of 0.95 and 0.96. MCSIMEX* has slightly smaller absolute bias. Because the analysis uses pooled estimates of $(\Pi,\bp)$, this design evaluates the pooled-nuisance approximation; the binary analysis also uses the nonidentity-link approximation. Complete results based on $N=500$ Monte Carlo replicates are reported in Table~A12 in Appendix~5. The corresponding BMI and obesity analyses in the UK Biobank follow in Section~\ref{sec:data analysis}.

\section{Application to UK Biobank data}\label{sec:data analysis}
In this application, we examine the associations of categorized daily fat intake with body mass index (BMI) and obesity in the UK Biobank. For both outcomes, the target is the category contrast $\theta_5-\theta_1$, which compares the highest and lowest quintiles of daily fat intake. The analysis includes $4{,}778$ white British participants ($2{,}324$ males and $2{,}454$ females) who completed two online 24-hour dietary recalls, the first between June and September 2011 and the second between October and December 2011. We use daily fat intake from the first recall as the error-prone exposure measurement $W$ and categorize it into quintiles as $\bW^c$ for fitting both outcome models. The two recalls are treated as repeated measurements and used to estimate the measurement error variance $\sigma_u^2$ under Assumption~\ref{assum1} and eq~\eqref{eq:equations_sigma_u}; the resulting $\Pi$ is defined for the categorized first recall. We fit a linear regression model for BMI and a logistic regression model for obesity, defined as BMI of at least 30, with age and sex included as error-free covariates. Age is centered, and sex is coded as 1 for male and 0 for female. A data summary and the fitted age and sex coefficients are reported in Appendix~6 and Table~A15.

\subsection{Assumption checks and diagnostics}
We first assess the measurement error model used to estimate $\Pi$ and $\bp$. Following \citet{eckert1997transformations} and \citet[Chapter~1.7]{carroll2006nonlinear}, we examine skewness, kurtosis, and normal QQ plots on the original and transformed scales. The average observed fat intake across the two recalls is right-skewed, with skewness 0.92 and kurtosis 4.95. With the estimated Box-Cox parameter $\widehat{\lambda}=0.3434$, the average of the Box-Cox transformed recalls has skewness 0.04 and kurtosis 3.45. Its QQ plot agrees closely with a normal distribution over the central range, with some remaining tail deviation (Figure~\ref{fig:qqplot_fat}, middle panel). The difference between the Box-Cox transformed recall measurements is also approximately Gaussian (Figure~\ref{fig:qqplot_fat}, right panel). Its sample mean is 0.07, and a $t$-test gives a $p$-value of 0.62, providing no evidence of a systematic mean shift between recall occasions. Regressing the transformed replicate difference on $\bZ$, which includes age and sex, gives small coefficients that are not statistically significant (Table~A16, Appendix~6). 
Together, these diagnostics are consistent with the transformed-scale replicate-error representation used in the analysis.

\begin{figure}
    \centering
    \includegraphics[scale=0.55]{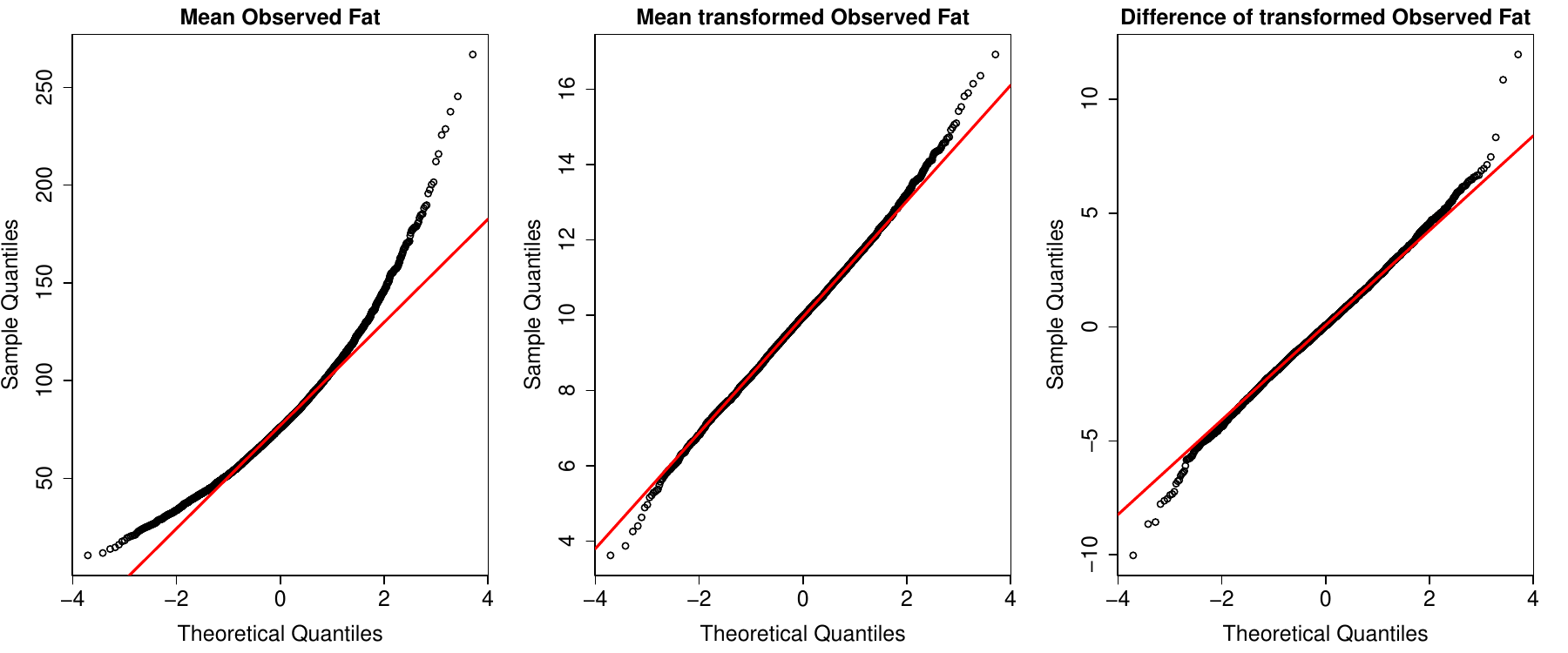}
    \caption{QQ plots for daily fat intake in the UK Biobank data. Left: average observed intake over two recalls. Middle: average Box-Cox transformed intake over two recalls. Right: difference between the Box-Cox transformed recall measurements.}
    \label{fig:qqplot_fat}
\end{figure}

We next examine the nuisance quantities used in the covariate-adjusted analysis. The correlation of $W$ with age $Z_2$ is $-0.02$, whereas its correlation with sex $Z_1$ is 0.16. We therefore compare the pooled nuisance estimates with estimates obtained separately for males and females. With $Z_1=1$ for males and $Z_1=0$ for females, we estimate $\widehat\Pi(Z_1=1)$, $\widehat\Pi(Z_1=0)$, $\widehat\bp(Z_1=1)$, and $\widehat\bp(Z_1=0)$. Relative to $\widehat\Pi$ and $\widehat\bp$, the maximum absolute deviations are less than 0.02 for both sex-specific misclassification matrices and less than 0.03 for both sex-specific category probability vectors. Thus, the sex-specific and pooled nuisance estimates are similar in these data. The linear BMI analysis uses the identity-link formulation, and the logistic obesity analysis uses the covariate-adjusted nonidentity-link approximation. Section~\ref{subsec:simu_additional} reports finite-sample results in settings with additional covariates, including a setting in which $X$ is correlated with $\bZ$.

\subsection{Results}\label{subsec:dataresults}
The estimated misclassification matrix for $J=5$, obtained using Algorithm~\ref{algo:pi_p}, is shown in the left panel of Figure~\ref{fig:intro}. We compare SIMFEX with the \emph{naive} analysis, which categorizes the error-prone $W$ as if it were $X$, and with MCSIMEX*, which uses the misclassification matrix estimated by Algorithm~\ref{algo:pi_p}. For BMI, the SIMFEX estimate of $\theta_5-\theta_1$ is 0.535, more than 40\% larger than the \emph{naive} estimate of 0.380; the MCSIMEX* estimate is 0.612 (Table~\ref{tab:dataresults}). The SIMFEX standard error is 0.266, compared with 0.194 for the \emph{naive} analysis, and its 95\% confidence interval is $(0.015,1.056)$ with a $p$-value of 0.044. Thus, the SIMFEX analysis gives a larger estimated BMI contrast together with greater estimated uncertainty than the \emph{naive} analysis.

\begin{table}
\begin{center}
\caption{Results for the high-versus-low fat-intake category contrast in the UK Biobank analysis.}\label{tab:dataresults}
\resizebox{\columnwidth}{!}{
\begin{tabular}{llcccc}
  \toprule
  Model & Method & Estimate & SE & 95\% CI & $p$-value\\ 
  \midrule
  linear
    & \emph{naive} & 0.380 & 0.194 & (0.000, 0.760) & 0.050 \\ 
    & MCSIMEX*     & 0.612 & 0.274 & (0.076, 1.149) & 0.025 \\ 
    & SIMFEX       & 0.535 & 0.266 & (0.015, 1.056) & 0.044 \\   
  \midrule
  logistic
    &\emph{naive} & 0.193 & 0.122 & (-0.045, 0.432) & 0.112\\
    & MCSIMEX*    & 0.251 & 0.179 & (-0.101, 0.602) & 0.163\\
    & SIMFEX      & 0.284 & 0.167 & (-0.045, 0.612) & 0.090 \\ 
  \bottomrule
\end{tabular}
}
\end{center}
\end{table}

For obesity, fitted by a logistic model with link function $g(p)=\ln\{p/(1-p)\}$, the SIMFEX estimate of $\theta_5-\theta_1$ is 0.284, compared with 0.193 for the \emph{naive} analysis and 0.251 for MCSIMEX* (Table~\ref{tab:dataresults}). The SIMFEX estimate is therefore larger than the \emph{naive} estimate, while its 95\% confidence interval $(-0.045,0.612)$ includes zero and its $p$-value is 0.090.

Overall, SIMFEX gives larger estimated contrasts between the highest and lowest categories than the \emph{naive} analysis for both BMI and obesity. The 95\% confidence interval for the BMI contrast excludes zero, whereas the interval for the obesity contrast includes zero. The application illustrates how measurement error adjustment can change both the estimated magnitude and uncertainty of associations based on a categorized continuous exposure.

\section{Discussion}\label{sec:conclusion}
The mean-scale formulation of SIMFEX separates estimation of the misclassification quantities from downstream outcome analysis. Under the stated conditions, it maps target means for latent exposure categories to observed-category outcome means. In the baseline setting without additional error-free covariates, this formulation gives a common construction for known one-to-one links and replaces MCSIMEX* pseudo-data generation and repeated outcome-model fitting with deterministic calculations.

The numerical studies show the value of assessing bias, RMSE, and interval calibration together. Across the main settings, SIMFEX substantially reduced absolute bias relative to the \emph{naive} analysis. In the primary $J=5$ study, it had slightly larger absolute bias than MCSIMEX* but smaller RMSE and coverage closer to nominal; in the simpler $J=3$ study, their point-estimation performance was comparable, and SIMFEX coverage remained close to nominal. The UK Biobank analysis further shows that accounting for category misclassification can materially alter both the estimated high-versus-low contrast for latent fat-intake categories and its uncertainty.

The formal results also clarify how the construction extends to covariate-adjusted analyses. For identity-link additive models, the contrast relation remains exact under the stated conditions, including covariate-invariant $\Pi$ and $\bp$. For nonidentity links, the same deterministic calculation provides a practical approximation because mixtures over latent categories need not remain in the corresponding generalized linear model. When $\Pi(\bz)$ or $\bp(\bz)$ varies with $\bz$, using pooled nuisance quantities introduces a distinct approximation. Thus, contrasts obtained by categorizing an error-prone measurement should be interpreted as observed-category quantities unless the induced category misclassification is addressed.

\endgroup

\section*{Acknowledgements} 
The authors thank the Editor, Associate Editor, and reviewers for insightful and constructive comments, which have greatly helped improve the paper. This research has been conducted using the UK Biobank Resource under Application Number 207159.

\section*{Supplementary material}
The appendix includes additional literature review, mathematical details on model identifiability, the SIMFEX method for bivariate contaminated covariates, proofs, sensitivity analyses, additional simulation results with covariates, an application-motivated simulation, and supplementary figures and tables for the simulation studies and UK Biobank analysis.

\bibliographystyle{abbrvnat}
\bibliography{R2-ref}

\end{document}